\documentclass{vgtc}                             

\usepackage{hyperref}
\usepackage{nameref}
\usepackage{cleveref}

\makeatletter
\let\orgdescriptionlabel\descriptionlabel
\renewcommand*{\descriptionlabel}[1]{%
  \let\orglabel\label
  \let\label\@gobble
  \phantomsection
  \edef\@currentlabel{#1}%
  \let\label\orglabel
  \orgdescriptionlabel{#1}%
}
\makeatother

\ifpdf
  \pdfoutput=1\relax                   
  \usepackage{graphicx}                
  \DeclareGraphicsExtensions{.pdf,.png,.jpg,.jpeg} 
\else
  \usepackage{graphicx}                
  \DeclareGraphicsExtensions{.eps}     
\fi%

\graphicspath{{figures/}{pictures/}{images/}{./}} 
\usepackage{comment}
\usepackage{microtype}                 
\PassOptionsToPackage{warn}{textcomp}  
\usepackage{textcomp}                  
\usepackage{mathptmx}                  
\usepackage{times}                     
\usepackage{cite}                      
\usepackage{tabu}                      
\usepackage{booktabs}                  
\usepackage{enumitem}

\usepackage[dvipsnames]{xcolor}
\usepackage[normalem]{ulem}

\onlineid{0}

\vgtccategory{Research}

\vgtcinsertpkg

\title{ElectroLens: Understanding Atomistic Simulations through Spatially-Resolved Visualization of High-Dimensional Features}

\author{Xiangyun Lei\thanks{e-mail: \{xlei38 $\mid$ fredhohman $\mid$ polo $\mid$ ajm\}@gatech.edu}\\ %
\and Fred Hohman\footnotemark[1]\\ %
\and Duen Horng Chau\footnotemark[1]\\ %
\and Andrew J. Medford\footnotemark[1]}

\teaser{
\centering
\includegraphics[width=0.93\textwidth]{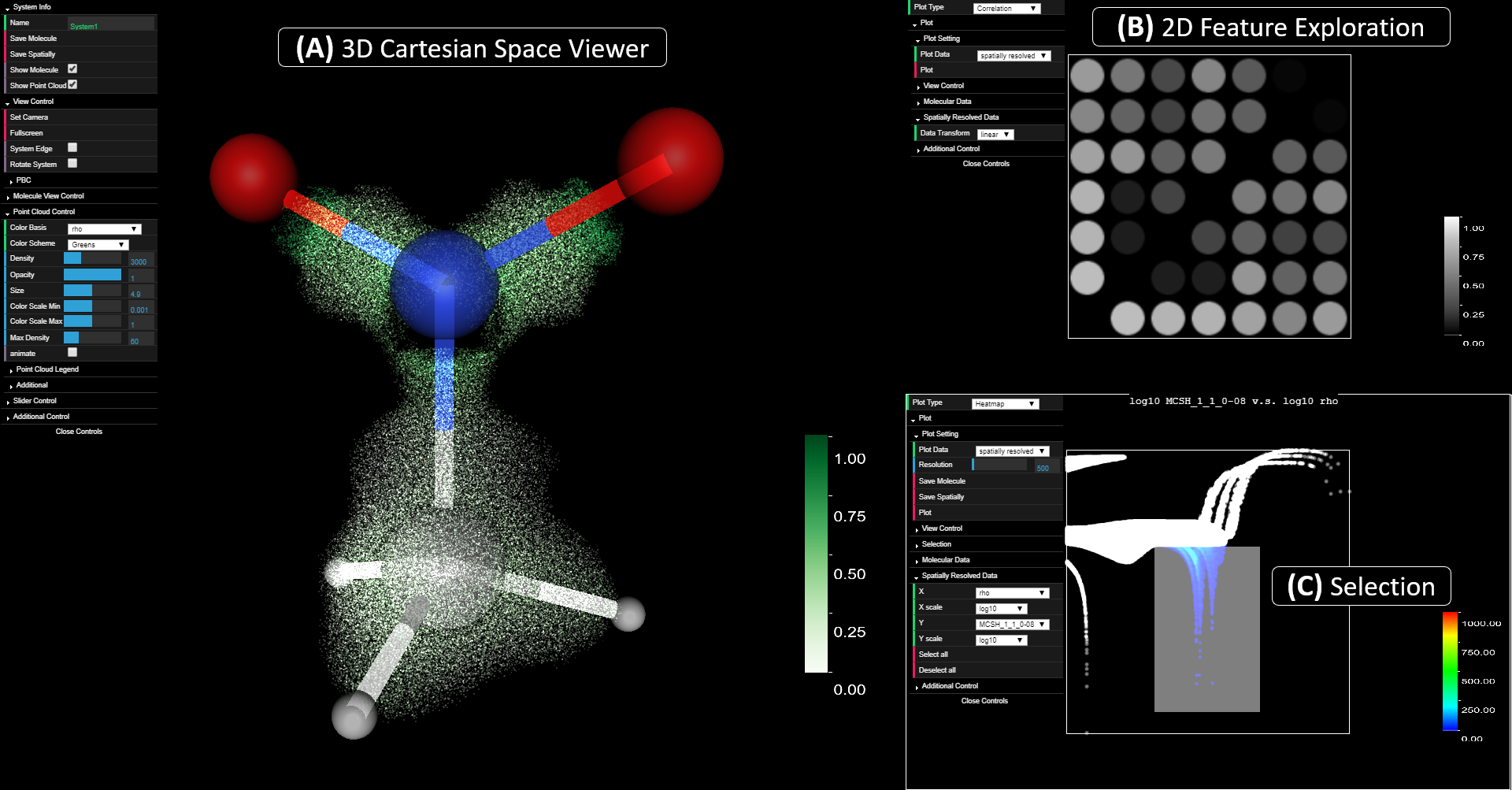}
\caption{The ElectroLens user interface (UI) visualizing the electron cloud of $CH_3NO_2$ molecule. 
The UI has two main parts: 3D viewer(s) on the left, and 2D plots on the right with option boxes in each view for customization. 
\textbf{(A)} The 3D viewer for Cartesian space. ElectroLens uses a ``point cloud'' to mimic the electron cloud, where the density of points corresponds to the density of electron cloud, and the color corresponds to energy density in this case. ElectroLens utilizes the ball-and-stick model to visualize atoms, where different atom types are denoted by white (hydrogen), grey (carbon), blue (nitrogen), and red (oxygen) spheres.  
\textbf{(B)} 2D plots for exploring and plotting additional features. The upper plot is the correlation plot used to identify the potentially interesting combinations of features. The plot below is a scatter plot of two features (in log-scale), where the color codes the number of data points represented.
\textbf{(C)} Selecting parts on the scatter plot causes the corresponding points in 3D viewer and other scatter plots to be highlighted. This case illustrates the connection between the features (right) and the chemical concept of a C-N bond (left).}
\label{fig:TeaserImage}
}

\abstract{In recent years, machine learning (ML) has gained significant popularity in the field of chemical informatics and electronic structure theory. These techniques often require researchers to engineer abstract ``features'' that encode chemical concepts into a mathematical form compatible with the input to machine-learning models. However, there is no existing tool to connect these abstract features back to the actual chemical system, making it difficult to diagnose failures and to build intuition about the meaning of the features. We present ElectroLens, a new visualization tool for high-dimensional spatially-resolved features to tackle this problem. The tool visualizes high-dimensional data sets for atomistic and electron environment features by a series of linked 3D views and 2D plots. The tool is able to connect different derived features and their corresponding regions in 3D via interactive selection. It is built to be scalable, and integrate with existing infrastructure.
} 

\CCScatlist{
  \CCScatTwelve{Molecular Visualization}{Visual Design}{Co\-ordinat\-ed and Multi\-ple Views}{Interaction Design}
}

\begin{document}
\maketitle
\section{Introduction}
\label{sec:Introuction}

Machine learning (ML) has seen lots of successful applications in the fields of chemical informatics and electronic structure theory in recent years\cite{Botu2019,Kitchin2018}. Scientists have applied various fingerprinting approaches to describe atoms or their associated electronic environments in molecular systems, and trained ML models that connects these high-dimensional feature vectors to some properties of interest.

One challenge that emerged from these studies is the lack of connection between these ML inputs and the underlying molecular systems. 
More specifically, with a set of feature vectors, there are no current tools for scientists to easily locate the corresponding atoms or electronic environments in the molecular systems. 
This is an issue for chemists, who typically develop strong ``chemical intuition'' through years of research and development experience. 
This intuition is based on the location and context of atoms or electrons in molecular systems, not on the derived mathematical features of ML models. 
The lack of connection prevents chemists from understanding the features intuitively. This makes it difficult for them to deduce the reason behind unsuccessful fingerprinting systems, or to improve them based on their chemical knowledge. 
Currently, researchers work around this mostly by trial and error: keep trying new fingerprinting system until the accuracy of the model provides indirect evidence that it works. While some systematic approaches have emerged \cite{Imbalzano_2018}, they are still based on enumeration of many possible candidates and do not enable chemists to apply their domain knowledge to select or analyze the resulting fingerprints. 

For example, a researcher might want to select features that differentiate the electronic structure of a bond between carbon (C) and nitrogen (N). This could be achieved by analyzing the electronic structure of a CH$_3$NO$_2$ molecule as shown on the left of Fig. \ref{fig:TeaserImage} along with features derived using convolutional kernels \cite{MLXC2019} (right). By identifying and selecting sub-sets of these features the researcher can learn that when the combination of two features is within certain window (shown by the selection in the bottom right) the C-N bond is selected. This enables the researcher to identify this as a critical fingerprint in subsequent machine-learning models.

Visualization has strong potential to solve this problem, enabling users to not only look at the high-dimensional feature vectors, but also establish the connections between them and their corresponding regions in the actual system. 
The connection will help researchers link the features to their chemical domain knowledge for easier understanding. 
This is highlighted by the example above, where a researcher can use visualization to establish an connection between the numerical range of certain features and chemical concepts like C-N bonding, which can facilitate feature engineering and model improvement. 
Moreover, by considering the prediction accuracy of a ML model as an additional feature, the same tool can be used to directly visualize the performance of the model, identify problematic regions, and improve the model accordingly.

\textbf{Contributions.} In this study, we work with experts in ML models for electronic structure theory and contribute:
\begin{itemize}[topsep=0pt,itemsep=0ex,partopsep=0ex,parsep=0.0ex]
    \item \textbf{ElectroLens, a 3D visualization tool} for high-dimensional spatially-resolved features associated with atoms and electronic environments of molecular systems. The UI of ElectroLens, shown in  \autoref{fig:TeaserImage}, consists of two parts: the 3D view(s) (left) corresponding to Cartesian space, and 2D plots (right) corresponding to projections of the feature space.
    
    \item \textbf{Interactive visualization design} that displays high dimensional data through a series of 3D and 2D plots and connects them via interactive selection (\autoref{sec:ElectroLens}). ElectroLens allows the user to make selections on the 2D plots, and the corresponding regions in real space will be highlighted in the 3D view. 
    
    \item \textbf{Scalable implementation} that can process and visualize more than \textit{one million} data points at a high frame rate of 60 FPS, on a commodity laptop computer.
    
    \item \textbf{An open-source desktop application with Python bindings} to the Atomic Simulation Environment (ASE) library \cite{ase-paper} that is commonly used by scientists for managing atomistic simulations. ElectroLens is currently hosted on Github at \url{medford-group.github.io/ElectroLens/}. Tutorials and documentation are provided on the website.
    
    \item \textbf{Usage scenarios} illustrate how ElectroLens helped to decipher new descriptor systems and ML model performance in our research on ML models for electronic and atomistic structures.
\end{itemize}

\section{Background}
\label{sec:Background}
Fingerprinting approaches are used to construct descriptive features to capture different aspects of molecular systems in the communities of ML electronic structure theory and chemical informatics. One common approach is to fingerprint molecular structures, leading to atom-centered features that capture the local chemical environment of a specific atom. There are numerous schemes including overlap integrals of Gaussians or atomic orbitals \cite{BehlerParrinello,BartokSOAP2013}, Zernike polynomials \cite{KHORSHIDI2016310}, and others \cite{Chandrasekaran2019}. The result of these schemes is a high (typically $>$10) dimensional vector that describes each atom, which can be described as vectors corresponding to points on a irregular grid.
An alternative approach is to examine the electronic structure surrounding the atoms instead of the atoms themselves. One strategy to achieve this is to treat the electron cloud as a voxelized 3D image, and to construct features to describe each voxel. For example, the recently-developed MCSH descriptors work in this way \cite{MLXC2019}.
In this case, the resulting feature vectors correspond to points on a regular grid. In both cases the problem involves high-dimensional features that correspond to points in 3D Cartesian space.

The existing fingerprinting systems have been the foundation of numerous successful ML models for predicting the energies from atomic configurations \cite{Botu2019, Behler2016}, or energy contributions from local electronic structure \cite{MLXC2019}. However, there is a significant challenge in understanding the physical or chemical meaning of the features which are typically based on mathematical transformations rather than physical derivations. This problem will likely increase with the rise of deep learning techniques where the features are determined by the algorithm, making it even harder to assign specific meaning. Visualization provides a promising route to gain intuition about big, high-dimensional data sets and corresponding ML models \cite{7784854,Hohman2018}.
While there are numerous software packages available for visualizing atomistic data sets \cite{HUMP96,VESTA32011,MoldenPaper1,OpenStructure, MView,Jmol,Pymol,Tomviz}, they have all been developed with chemical properties in mind, and are optimized for visualizing chemical concepts such as atom types, chemical bonds, and electron density. However, none are optimized to visualize the high-dimensional features used in ML models. Most programs are limited to visualizing a maximum of 2 features other than position (typically element type and radius) with a single 3D view, and often become sluggish when visualizing more than ten thousand data points. The data sizes required for ML models typically exceed this limit in terms of number of data points and dimensions per point.

\section{Design Challenge}
\label{sec:DesignChallenge}

To establish an intuitive link between feature vectors and actual systems, we have worked with domain experts to identify six design challenges that are currently not addressed by other tools in the electronic structure theory or chemical informatics communities. 

\begin{description}[topsep=0pt,itemsep=0ex,partopsep=0ex,parsep=0.0ex]
    \item[C1\label{dc:highdim}] \textbf{Visualizing high-dimensional feature vectors.} The typical fingerprinting systems used in research generates tens or even hundreds of features per atom or grid point. Visualization of the distribution and correlations between these features is crucial to the understanding of their meaning.

    \item[C2\label{dc:connecting}] \textbf{Connecting features and corresponding Cartesian space.} It is critical to visualize the features in their corresponding chemical context. A seamless connection between these alternative representations of the system is important to grasp the meanings of the features. Unfortunately, there is no existing tool in the community that establishes these connections visually.

    \item[C3\label{dc:systems}] \textbf{Comparing features across different systems.} Assessing the generality of a given relationship between a feature vector and a chemical system requires comparison of multiple systems simultaneously. For example, if a feature is found to correspond to the C=O bonding region in a CO$_2$ molecule, it would be of interest to know if the same feature also corresponds to the C=O bond in a similar molecule such as HCOOH. However, most existing visualization tools for atomic/electronic structure focus on visualizing a single system at a time.
    
    \item[C4\label{dc:structures}] \textbf{Visualizing atomistic and electronic structures simultaneously.} The atomic structure and electronic structure are intimately linked, but are represented with different data structures: irregular grids for atoms, and regular grids for electronic environments. The ability to simultaneously visualize the two data types, along with features that describe environments within these two related structures, will provide a new route to building intuitive connections between the two representations.

    \item[C5\label{dc:scale}] \textbf{Handling large datasets.} Datasets used for ML training are often very large, containing millions of data points with tens of dimensions or more. Most tools for visualizing atomistic and electronic structure data do not scale well to datasets of extreme size. Building intuition requires interactive visualization of such large datasets, so fast rendering speed is critical.

    \item[C6\label{dc:infrastructure}] \textbf{Integrating with existing infrastructure.} Simulations of atomistic and electronic structure involve a wide range of tools and software packages, including molecular dynamics, density functional theory, and wavefunction theory codes. These tools have a range of input/output data structures, so it is important for visualization tools to integrate with existing software packages that already act as API's for integrating the tools.

\end{description}

\begin{figure}
\centering
\includegraphics[width=0.9\columnwidth]{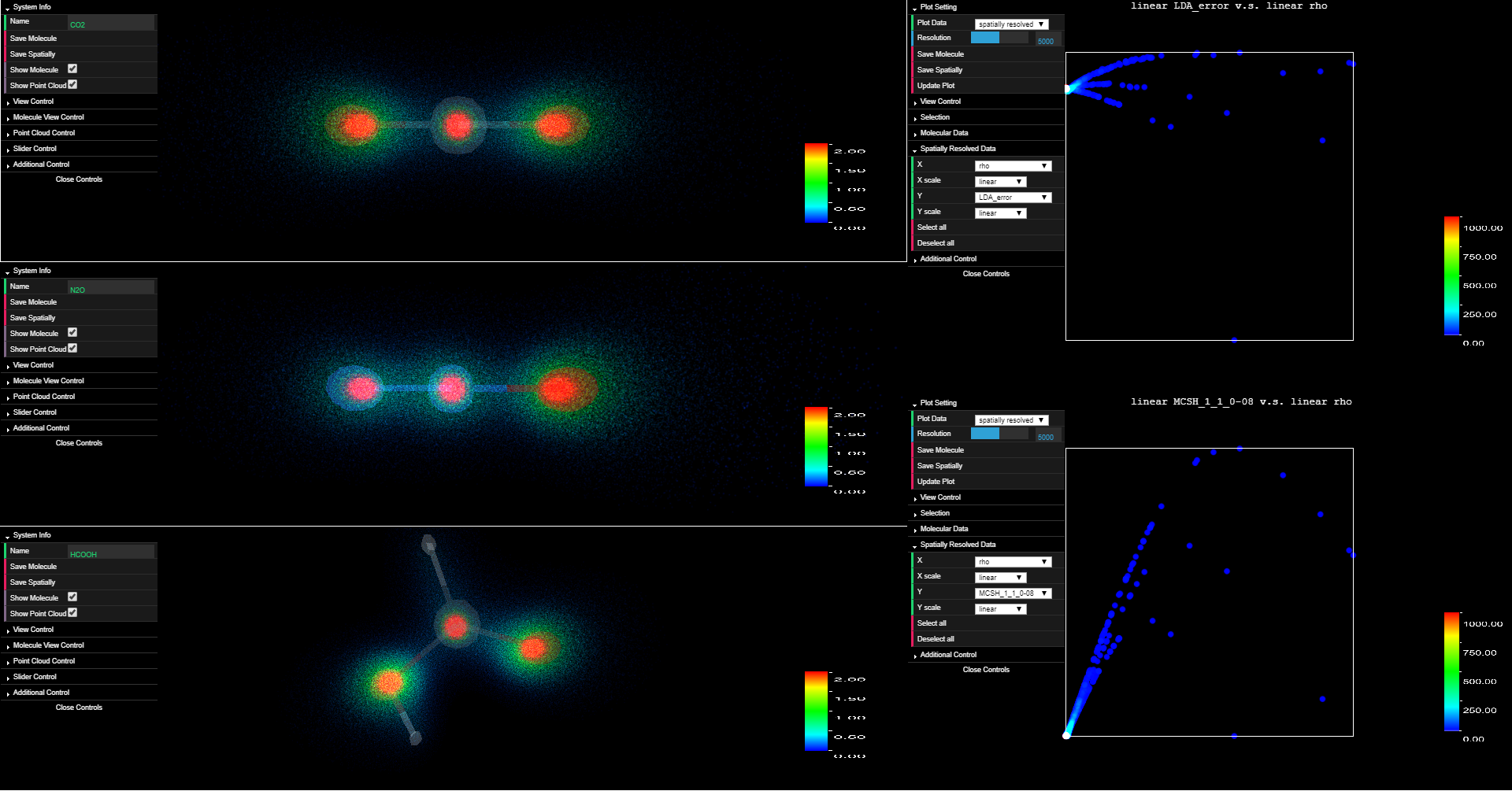}
\setlength{\belowcaptionskip}{-4pt}
\caption{ElectroLens simultaneous viewing atomistic (spheres) and electronic environments (point cloud) for multiple systems. The chemical systems shown are carbon dioxide (CO$_2$) (top), nitrous oxide (N$_2$O) (middle), and formic acid (HCOOH) (bottom). The 2D plots show the error of an ML model vs. electron density (top) and the first derivative of the electron density vs. electron density (bottom). Features in the 2D plots correspond to all 3 systems.}
\label{fig:multipleSystems}
\end{figure}

\begin{figure}
\centering
\includegraphics[width=0.9\columnwidth]{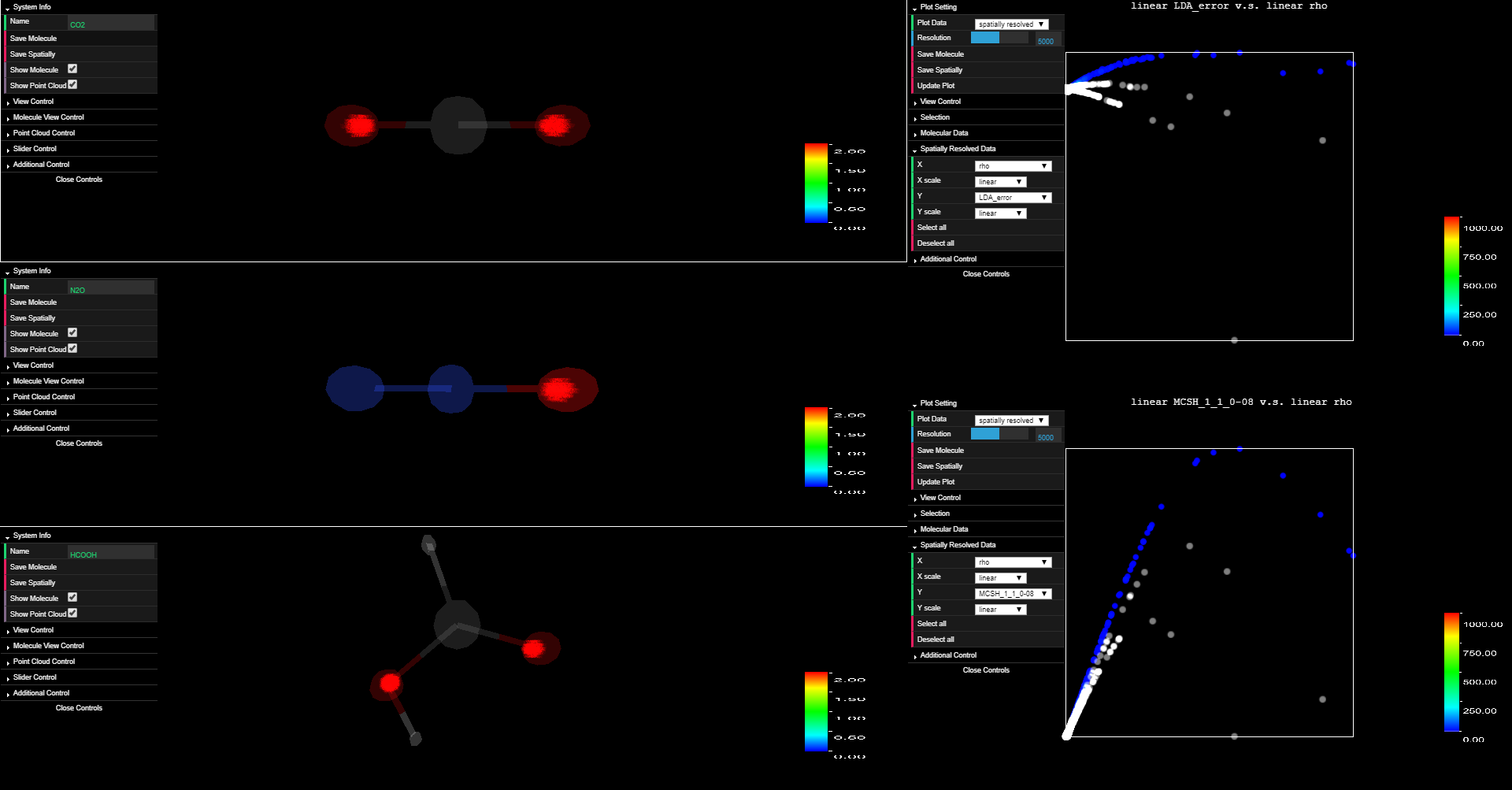}
\setlength{\belowcaptionskip}{-9pt}
\caption{The same plots as Fig. \ref{fig:multipleSystems} with selection. The top ``tail'' of the error distribution vs. electron density plot is selected, causing the O core regions to be highlighted in the 3D views. The top ``tail'' of the derivative plot is also highlighted as a result, illustrating a connection between the error and the density derivative.}
\label{fig:multipleSystemsAfterSelection}
\end{figure}

\section{ElectroLens}
\label{sec:ElectroLens}
\subsection{Main Views and Interactions}
\label{sec:MainView}

ElectroLens is written in JavaScript with the \textit{three.js} library. The WebGL-based technology is lightweight and enables high frame-rate (60fps) rendering of millions of points with a GPU [C5]. It is wrapped into a standalone desktop app that supports Windows, MacOS and Linux with the \textit{Electron JS} framework \cite{Electronjs}. Memory usage is optimized by adapting an efficient data structure where each data point is only stored once and shared between multiple plots without redundant copies. ElectroLens is structured to have two main views: the 3D view on the left, and the 2D plots on the right (Fig. \ref{fig:TeaserImage}). These views are dynamically linked, allowing selections in the 2D plots to highlight corresponding regions in the 3D view.

\begin{figure}
\centering
\includegraphics[width=0.9\columnwidth]{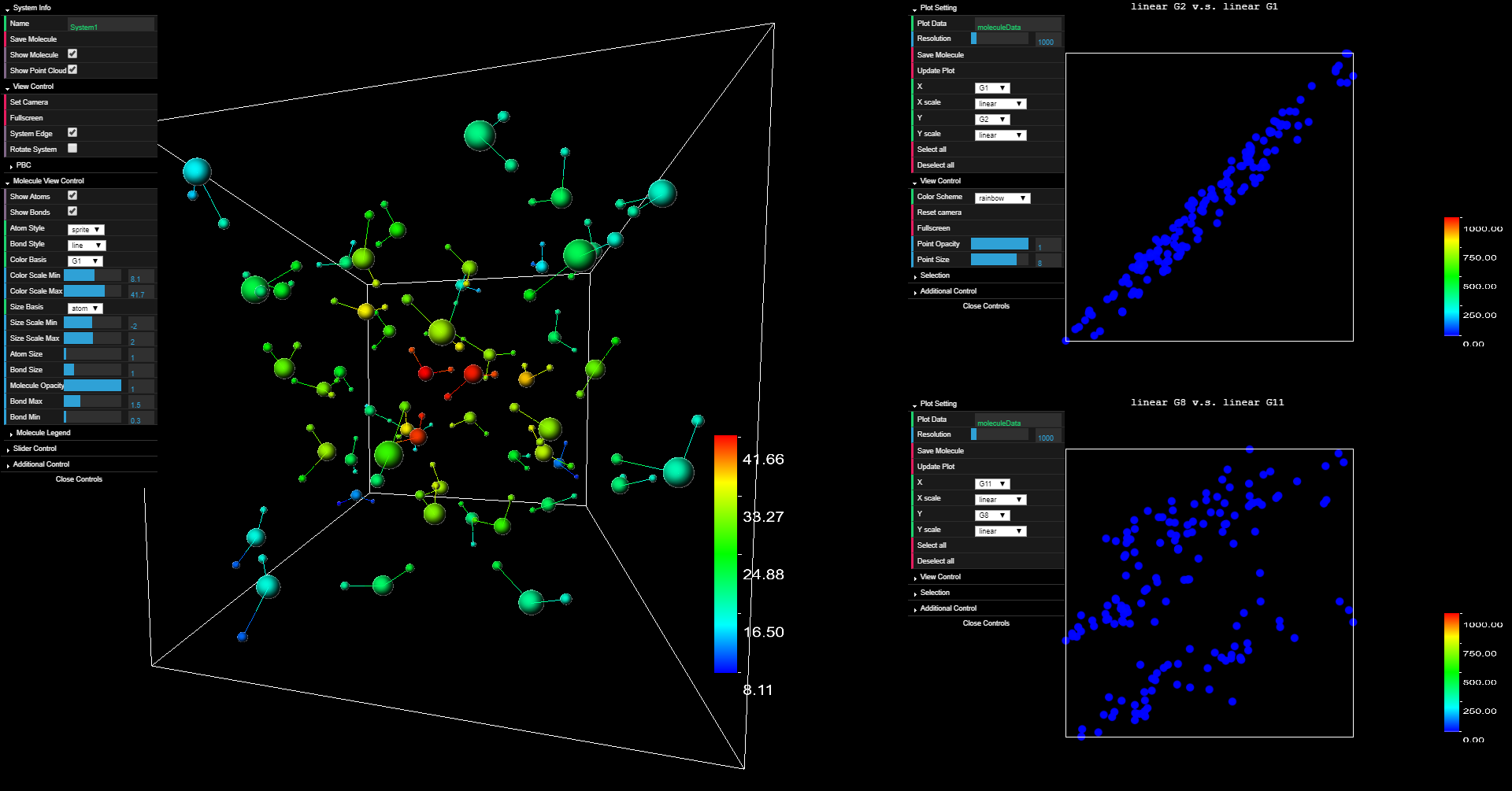}
\setlength{\belowcaptionskip}{-9pt}
\caption{ElectroLens visualizing a cluster of water (H$_2$O) molecules. For the 3D view, the size codes atom type (large = oxygen, small = hydrogen), and the color codes the values of a symmetry function fingerprint \cite{BehlerParrinello}. The lines between atoms label the bonds. Both 2D plots are plotting two symmetry function features, and users can select parts on the 2D plots to view the corresponding atoms in the 3D view.}
\label{fig:multipleDescs}
\end{figure}

\textbf{3D View.} The 3D view corresponds to physical systems in Cartesian space. For atomistic simulations, the physical information can take two forms: atom positions (irregular grids) or electron/wavefunction densities (regular grids). ElectroLens can visualize these distinct data types separately or simultaneously, as shown in Figs. \ref{fig:multipleSystems}-\ref{fig:multipleDescs}. For the atomic features, ElectroLens uses the common ``ball-and-stick'' model to display the location of the atoms and bonds. The size and color of the ``balls'' can be used to code 2 features of interest, which are assumed to be the atomic type by default, but could be other features. For the electronic environment features, ElectroLens renders this data as a point cloud to mimic the electron cloud, where the density of points corresponds to the density of electron cloud, and the color can be used to encode an extra feature. The view is highly customizable for users to adjust visualization and highlight regions of interest. For example, users can adjust the size, transparency, color scale, color map, or density of the point cloud with the option box of the view. On the other hand, one could also change the resolution, color, size of the ball-and-stick model and more as well. The view also supports ``slicing'', allowing users to look at cross sections of the electron density.

\textbf{2D Plots and Selection.} The 3D view is limited to visualizing 2 spatially-resolved features through double-encoding on size/color (ball and stick) or density/color (point clouds). To display additional features without loss of information 
[C1], a series of connected 2D plots are used. 
Three kinds of 2D plots are implemented: scatter plots, correlation plots and dimension-reduction plots.
The scatter plots are heat maps where two features of choice are plotted and the color of the heat map corresponds to the amount of points at a given region of feature space. 
The correlation plot is a visual representation of the correlation matrix of the features. This provides a fast way to explore the high-dimensional feature space and identify interesting combinations, since feature combinations with low correlation tend to contain the most information. The dimension-reduction plots provide an alternate strategy for navigating the high-dimensional space. They utilize principal component analyis (PCA) reduce the dimensionality of the features and plot the resulting projection using a scatter plot. 
These 2D plots can be added to the right-hand view on the fly with common transformations like ``log10'' for better visualization. Users can also choose to simultaneously plot features corresponding to atomic and electronic environments to assess connections between the two 
[C4]. One key functionality of the scatter plots and dimension-reduction plots is the ability to select regions in the 2D plots and see how these features are localized in 3D Cartesian space 
[C2], as highlighted in Fig. \ref{fig:TeaserImage}. ElectroLens supports exporting these sub-sets of points for further analysis, as well as saving specific views that can be loaded in later or shared with other users.
Users can customize the 2D plots (color map, scatter size, etc.) as well.

\textbf{Multiple 3D System Views.} 
ElectroLens allows user to visualize and compare an arbitrary number of systems simultaneously, and use the 2D plots to plot the pooled data across all systems (Fig. \ref{fig:multipleSystems}) to understand the generality of connections between feature vectors and chemical concepts [\ref{dc:systems}]. In other words, points in the 2D plots may correspond to regions in multiple systems. When the points are selected in the 2D view, the corresponding regions in all systems where the points appear will be highlighted, as illustrated in Fig. \ref{fig:multipleSystemsAfterSelection}.


\textbf{Python and ASE bindings.} 
Many researchers who conduct atomistic simulations use Python to setup simulations and analyze results. Therefore, we have also created Python bindings for ElectroLens by wrapping with the \texttt{CEFpython} library \cite{Tomczak2019cefpython}. The bindings are designed to follow the structure of the ASE Python library \cite{ase-paper}, which is widely used by researchers in the field and contains classes corresponding to common data structures like atom positions [\ref{dc:infrastructure}]. 
ASE also contains numerous translators for reading/writing a range of different file types into standardized data structures. By leveraging Python and ASE, ElectroLens is compatible with a wide range of file formats, and easy to use for anyone familiar with ASE.

\subsection{Case Studies}
\label{sec:UsageScenario}
This section presents two example applications of ElectroLens, presenting how it facilitated and accelerated research in ML and chemistry. They are: (1) Diagnosing the failure of a neural network for predicting exchange-correlation energy from electron density (Sect. \ref{sec:scenario1}) and (2) Identifying atomic configurations corresponding to a failing machine-learned force field (Sect. \ref{sec:scenario2}).

\subsubsection{Constructing ML models from electron density}
\label{sec:scenario1}

ElectroLens was originally developed to support a research project to establish a ML model to predict the exchange-correlation energy, a key quantity needed in density functional theory (DFT) \cite{MLXC2019}. Briefly, DFT is a widely-used technique for simulating the electronic structure of molecular systems. The formalism is based on a powerful theorem proving a one-to-one mapping between electron density and ground-state energy \cite{doi:10.1063/1.4869598,RevModPhys.87.897, PhysRev.136.B864}. However, the connection to one part of the energy, known as the exchange-correlation (xc) energy, is unknown. Our work focused on using neural network (NN) to empirically learn this connection. As a first step, a NN model was trained between the local electron density at a point and the corresponding xc energy at the same point \cite{HandyAndTozer}. However, the resulting accuracy was insufficient, and ElectroLens was used to diagnose this result.

\textbf{Connecting chemistry to features.}  We diagnosed this problem by plotting the NN prediction error as a function of the electron density that was the input of the NN. This revealed a multi-valued error distribution with three distinct ``tails'' (Fig. \ref{fig:multipleSystems}, upper right). This explains the poor accuracy of the model because multi-valued functions can not be regressed, and implies there are electronic environments that can not be distinguished by electron density alone. However, the chemical nature of the problematic environments was not immediately clear. We selected several sample systems (CO$_2$, N$_2$O, and HCOOH) and utilized ElectroLens to simultaneously visualize the electronic environments and atom positions (Fig.\ref{fig:multipleSystems} left panels).  
Through interactive selection, ElectroLens revealed that the tails correspond to the atomic core regions for C, N, and O atoms, as shown in  Fig. \ref{fig:multipleSystemsAfterSelection}. This clearly illustrated that the NN model was failing due to the electron density in the atomic core regions.

\textbf{Understanding relationships between features.} To further diagnose the issue, additional features were selected to distinguish the C, N, and O core regions. Prior models utilize the derivative of the electron density, so this was one of the features we investigated. We used ElectroLens to plot the derivative as a function of the electron density and observed a similar three-tailed structure, as shown in Fig. \ref{fig:multipleSystems} (bottom right). ElectroLens was used to show that these tails corresponded to the tails in the error distribution (Fig. \ref{fig:multipleSystemsAfterSelection}). This provided the insight that the derivative of the electron density is capable of distinguishing the C, N, and O core regions, and led to a new and more accurate NN model based on the density and its derivative. This evolved into the development of new rotationally-invariant versions of the derivatives based on convolutions with MCSH kernels, ultimately improving the accuracy of the models even further \cite{MLXC2019}.

From this case study, two features have proven to be crucial: (1) The ability to simultaneously compare across different systems; (2) The ability to interact smoothly (60 fps) with a large (1 million 47-dimensional points) data set.

\subsubsection{Assessing ML force-fields based on atomic positions}
\label{sec:scenario2}
The construction of ML force-fields for predicting energies and forces directly from atomic structures is a rapidly growing sub-field of computational chemistry \cite{BehlerReview2015,Kitchin2018, Botu2019,Ying2017}. Briefly, these methods take atomic positions as inputs, then use symmetry-preserving fingerprints to distinguish the chemical environments experienced by each atom. These fingerprints are used as inputs to ML models, typically NNs, and trained to rapidly and accurately reproduce forces and energies of each atom using training data generated with expensive quantum-mechanical simulations. However, the black-box nature of the process makes it difficult to diagnose failures. In our research group we commonly experience a specific failure mode where a model is accurate for training data, but exhibits erroneously large forces when applied in a molecular dynamics simulation. This occurs because some atoms in the system move into chemical environments that were not sampled in the training data, leading to inaccurate results. However, it can be particulary difficult to pinpoint exactly which atoms are outliers, making it challenging to identify new training systems that will improve the model.

\textbf{View customization and double encoding.} 
ElectroLens supports encoding features or properties such as the force magnitude in the 3D visualization frame. This enables views like the one shown in Fig. \ref{fig:multipleDescs}, where atoms with excessively large forces can be easily identified, as they are colored by the extreme ends of the chosen color map. This can be used to construct new training data that contains atomic environments similar to those with large errors.

\textbf{Interactive selection with atomistic features.} 
In addition to visualizing errors in the 3D frame, it is also possible to plot them in the 2D frames, along with other input features to the model. This provides a route to identify regions of the feature space that have not been sampled, facilitating the generation of new training data or implementation of methods to raise errors when the model moves outside the domain of the training data.

The main lesson learned from this study is that the use of double-encoding features using color or size can assist researchers in identifying interesting patterns. This is not standard in visualizing chemical data, since size and color are typically determined by the element.

\section{Conclusion and Future Work}
\label{sec:Conclusion}

A new visualization tool called ElectroLens has been developed to analyze high-dimensional features derived from 3D datasets. The tool has been developed in the context of atomic and electronic structure data, corresponding to two different representations (irregular and regular grids). ElectroLens is built to be efficient with large datasets easy to use and integrate with existing infrastructure for atomstic simulations. ElectroLens was applied and tested in multiple scenarios, leading to improved ML models for exchange-correlation energies and atomistic force fields (Sect. \ref{sec:UsageScenario}). Future work includes implementing other projections into ElectroLens, as well as studying methods to automatically infer informative feature combinations. We also plan to focus on applications of the tool to the increasingly popular NN force-fields for molecular dynamics simulations \cite{BehlerReview2015,Kitchin2018, Botu2019,Ying2017}, including improved handling of time-dependent data sets. Further, we expect that ElectroLens may be useful for a wider set of problems in chemistry, physics, and engineering involving spatially-resolved high-dimensional data.

\acknowledgments{
This work is supported by the U.S. Department of Energy, Office of Science, Office of Basic Energy Sciences Computational Chemical Sciences program under Award Number DE-SC0019410, with support for atomistic feature visualization and integration with the AMPTorch code supported by Award Number DE-SC0019441. The authors are also grateful for a GPU generously provided by the NVIDIA Corporation through the GPU Grant program.}

\bibliographystyle{abbrv}

\bibliography{main}

\begin{thebibliography}{10}

\bibitem{Electronjs}
Electron.
\newblock \url{https://electronjs.org}.

\bibitem{Jmol}
Jmol: an open-source java viewer for chemical structures in 3d.
\newblock \url{http://www.jmol.org/}.

\bibitem{Tomviz}
Tomviz.
\newblock \url{https://tomviz.org/}.

\bibitem{BartokSOAP2013}
A.~P. Bart\'ok, R.~Kondor, and G.~Cs\'anyi.
\newblock On representing chemical environments.
\newblock {\em Phys. Rev. B}, 87:184115, May 2013.

\bibitem{doi:10.1063/1.4869598}
A.~D. Becke.
\newblock Perspective: Fifty years of density-functional theory in chemical
  physics.
\newblock {\em The Journal of Chemical Physics}, 140(18):18A301, 2014.

\bibitem{BehlerReview2015}
J.~Behler.
\newblock Constructing high-dimensional neural network potentials: A tutorial
  review.
\newblock {\em International Journal of Quantum Chemistry}, 115(16):1032--1050,
  2015.

\bibitem{Behler2016}
J.~Behler.
\newblock Perspective: Machine learning potentials for atomistic simulations.
\newblock {\em The Journal of Chemical Physics}, 145(17):170901, 2016.

\bibitem{BehlerParrinello}
J.~Behler and M.~Parrinello.
\newblock Generalized neural-network representation of high-dimensional
  potential-energy surfaces.
\newblock {\em Phys. Rev. Lett.}, 98:146401, Apr 2007.

\bibitem{OpenStructure}
M.~Biasini, T.~Schmidt, S.~Bienert, V.~Mariani, G.~Studer, J.~Haas, N.~Johner,
  A.~D. Schenk, A.~Philippsen, and T.~Schwede.
\newblock {{\it OpenStructure}: an integrated software framework for
  computational structural biology}.
\newblock {\em Acta Crystallographica Section D}, 69(5):701--709, May 2013.

\bibitem{Botu2019}
V.~Botu, R.~Batra, J.~Chapman, and R.~Ramprasad.
\newblock Machine learning force fields: Construction, validation, and outlook.
\newblock {\em The Journal of Physical Chemistry C}, 121(1):511--522, 2017.

\bibitem{Chandrasekaran2019}
A.~Chandrasekaran, D.~Kamal, R.~Batra, C.~Kim, L.~Chen, and R.~Ramprasad.
\newblock Solving the electronic structure problem with machine learning.
\newblock {\em npj Computational Materials}, 5:22, Feburary 2019.

\bibitem{PhysRev.136.B864}
P.~Hohenberg and W.~Kohn.
\newblock Inhomogeneous electron gas.
\newblock {\em Phys. Rev.}, 136:B864--B871, Nov 1964.

\bibitem{Hohman2018}
F.~Hohman, M.~Kahng, R.~Pienta, and D.~H. Chau.
\newblock Visual analytics in deep learning: An interrogative survey for the
  next frontiers.
\newblock {\em CoRR}, abs/1801.06889, 2018.

\bibitem{HUMP96}
W.~Humphrey, A.~Dalke, and K.~Schulten.
\newblock {VMD} -- {V}isual {M}olecular {D}ynamics.
\newblock {\em Journal of Molecular Graphics}, 14:33--38, 1996.

\bibitem{Imbalzano_2018}
G.~Imbalzano, A.~Anelli, D.~Giofr{\'{e}}, S.~Klees, J.~Behler, and M.~Ceriotti.
\newblock Automatic selection of atomic fingerprints and reference
  configurations for machine-learning potentials.
\newblock {\em The Journal of Chemical Physics}, 148(24):241730, jun 2018.

\bibitem{RevModPhys.87.897}
R.~O. Jones.
\newblock Density functional theory: Its origins, rise to prominence, and
  future.
\newblock {\em Rev. Mod. Phys.}, 87:897--923, Aug 2015.

\bibitem{KHORSHIDI2016310}
A.~Khorshidi and A.~A. Peterson.
\newblock Amp: A modular approach to machine learning in atomistic simulations.
\newblock {\em Computer Physics Communications}, 207:310 -- 324, 2016.

\bibitem{Kitchin2018}
J.~R. Kitchin.
\newblock Machine learning in catalysis.
\newblock {\em Nature Catalysis}, 1:230, Apr 2018.

\bibitem{ase-paper}
A.~H. Larsen, J.~J. Mortensen, J.~Blomqvist, I.~E. Castelli, R.~Christensen,
  M.~Dułak, J.~Friis, M.~N. Groves, B.~Hammer, C.~Hargus, E.~D. Hermes, P.~C.
  Jennings, P.~B. Jensen, J.~Kermode, J.~R. Kitchin, E.~L. Kolsbjerg, J.~Kubal,
  K.~Kaasbjerg, S.~Lysgaard, J.~B. Maronsson, T.~Maxson, T.~Olsen, L.~Pastewka,
  A.~Peterson, C.~Rostgaard, J.~Schiøtz, O.~Schütt, M.~Strange, K.~S.
  Thygesen, T.~Vegge, L.~Vilhelmsen, M.~Walter, Z.~Zeng, and K.~W. Jacobsen.
\newblock The atomic simulation environment—a python library for working with
  atoms.
\newblock {\em Journal of Physics: Condensed Matter}, 29(27):273002, 2017.

\bibitem{MLXC2019}
X.~Lei and A.~J. Medford.
\newblock Design and analysis of machine learning exchange-correlation
  functionals via rotationally invariant convolutional descriptors.
\newblock {\em Phys. Rev. Materials}, 3:063801, Jun 2019.

\bibitem{Ying2017}
Y.~Li, H.~Li, F.~C. Pickard, B.~Narayanan, F.~G. Sen, M.~K.~Y. Chan, S.~K.
  R.~S. Sankaranarayanan, B.~R. Brooks, and B.~Roux.
\newblock Machine learning force field parameters from ab initio data.
\newblock {\em Journal of Chemical Theory and Computation}, 13(9):4492--4503,
  2017.
\newblock PMID: 28800233.

\bibitem{7784854}
S.~{Liu}, D.~{Maljovec}, B.~{Wang}, P.~{Bremer}, and V.~{Pascucci}.
\newblock Visualizing high-dimensional data: Advances in the past decade.
\newblock {\em IEEE Transactions on Visualization and Computer Graphics},
  23(3):1249--1268, March 2017.

\bibitem{Pymol}
S.~LLC.
\newblock The pymol molecular graphics system.
\newblock \url{https://pymol.org/2/}.

\bibitem{VESTA32011}
K.~Momma and F.~Izumi.
\newblock {{\it VESTA3} for three-dimensional visualization of crystal,
  volumetric and morphology data}.
\newblock {\em Journal of Applied Crystallography}, 44(6):1272--1276, Dec 2011.

\bibitem{MoldenPaper1}
G.~Schaftenaar and J.~Noordik.
\newblock Molden: a pre- and post-processing program for molecular and
  electronic structures*.
\newblock {\em Journal of Computer-Aided Molecular Design}, 14(2):123--134, Feb
  2000.

\bibitem{Tomczak2019cefpython}
C.~Tomczak.
\newblock Keras.
\newblock \url{https://github.com/cztomczak/cefpython}, 2019.

\bibitem{HandyAndTozer}
D.~J. Tozer, V.~E. Ingamells, and N.~C. Handy.
\newblock Exchange‐correlation potentials.
\newblock {\em The Journal of Chemical Physics}, 105(20):9200--9213, 1996.

\bibitem{MView}
L.~Viani et~al.
\newblock Mview: A tool for visualization and analysis of molecular properties.
\newblock \url{http://mview-tools.com/}.

\end{thebibliography}
\end{document}